\documentstyle[12pt]{article}
\newcommand{\be}{\begin{equation}}
\newcommand{\ee}{\end{equation}}
\newcommand{\ba}{\begin{eqnarray}}
\newcommand{\ea}{\end{eqnarray}}

\begin{document}
\begin{center}
 {\bf\Large{
   Lagrangians with linear velocities within Riemann-Liouville fractional derivatives }}
\end{center}
\begin{center}
{\bf Dumitru Baleanu}\footnote[1]{ On leave of absence from
Institute of Space Sciences, P.O BOX, MG-23, R 76900
Magurele-Bucharest, Romania, E-mails: dumitru@cankaya.edu.tr,
baleanu@venus.nipne.ro}, {\bf Tansel
Avkar}\footnote[2]{E-Mail:~~avkar@cankaya.edu.tr}
\end{center}
\begin{center}
Department of Mathematics and Computer Science, Faculty of Arts
and Sciences, \c{C}ankaya University-06530, Ankara, Turkey
\end{center}
\begin{abstract}

Lagrangians linear in velocities were analyzed using the
fractional\\ calculus and the  Euler-Lagrange equations were
derived. Two examples  were investigated in details, the explicit
solutions of Euler-Lagrange equations were obtained and the
recovery of the  classical results  was discussed.

\end{abstract}

\section{Introduction}

    The mathematical idea of fractional derivatives, which
goes back to the seventeenth century, has represented the subject
of interest for various mathematicians \cite{Oldham, Miller,
Podlubny}. The fractional calculus, which means the calculus of
derivatives and integrals of any arbitrary real or complex order
is gaining a considerable importance in various branches of
science, engineering and finance \cite{Oldham, Miller, Podlubny,
Mainardi1, Mainardi2, hilfer, egipt,rosu,om2,engheta, finan}.

As it was observed during the past three decades, the fractional
differential calculus describes more accurately the physical
systems \cite{ Mainardi1,Mainardi2,hilfer,Bagley}.Many
applications of fractional calculus amount to replacing the time
derivative in an evolution equation with a derivative of
fractional order. This is not merely a phenomenological procedure
providing an additional fit parameter. One of the problems
encountered in this field is what kind of fractional derivatives
will  replace the integer derivative for a given problem
\cite{Oldham, Miller,Podlubny, Hilfer}. Depending on the specified
physical situation different authors have applied different
derivatives \cite{hilfer, om2}.

 Nonconservative Lagrangian
and  Hamiltonian mechanics were investigated by Riewe within
fractional calculus \cite{Riewe1,Riewe2}. Besides,
 Lagrangian and Hamiltonian fractional sequential mechanics,
 the models with symmetric fractional derivative were studied in
\cite{ klimek2,klimek1} and  the properties of fractional
differential forms were introduced \cite{naber, naber1}.

 Recently, an extension of the simplest fractional
variational problem and the fractional variational problem of
Lagrange was obtained  \cite{om1}. A natural and interesting
generalization of Agrawal's approach \cite{om1} is to apply the
fractional calculus to the constrained systems
\cite{const,baleanu2000}. In the present work, we apply the
concept of fractional calculus to the Lagrangian with linear
velocities.

The plan of this paper is as follows:\\
 In Sec. 2 the
Riemann-Liouville (RL) fractional derivatives were briefly presented.\\
 In Sec. 3 the fractional Lagrangians with linear velocities were
 constructed and their corresponded  Euler-Lagrange were analyzed. Sec. 4 was devoted to  our conclusions.

\section{Riemann - Liouville fractional derivatives}

In this section the definitions of the right and left RL
derivatives as well as their basic properties are briefly
presented.

 RL fractional derivatives
\cite{Podlubny, om1} are defined as follows
\begin{eqnarray}\label{unu}
{}_a
\textbf{D}_t^{\alpha}f(t)=\frac{1}{\Gamma{(n-\alpha)}}\left(\frac{d}{dt}\right)^{n}\int\limits_a^t(t-\tau)^{n-\alpha-1}f(\tau)d\tau
\ ,
\end{eqnarray}

\begin{eqnarray}\label{two}
{}_t
\textbf{D}_b^{\alpha}f(t)=\frac{1}{\Gamma{(n-\alpha)}}\left(-\frac{d}{dt}\right)^{n}\int\limits_t^b(\tau-t)^{n-\alpha-1}f(\tau)d\tau,
\end{eqnarray}
where the order $\alpha$ fulfills $n-1\leq\alpha <n$ and $\Gamma$
represents the Euler's gamma function. The first derivative  is
called the RL left fractional derivative  and in (\ref{two}) the
expression of  the RL right fractional is presented. If $\alpha$
is any integer, the relations to the usual derivative are obtained
as follows
\begin{eqnarray}
{}_a \textbf{D}_t^{\alpha}f(t)=\left(\frac{d}{dt}\right)^{\alpha},
\quad {}_t
\textbf{D}_b^{\alpha}f(t)=\left(-\frac{d}{dt}\right)^{\alpha}.
\end{eqnarray}

 Under the assumptions that $f(t)$ is continuous and  $p\geq q\geq
0 \ $, the most general property of RL fractional derivatives can
be written as
\begin{eqnarray}
{}_a \textbf{D}_t^{p}\left({}_a \textbf{D}_t^{-q}f(t)\right)={}_a
\textbf{D}_t^{p-q}f(t).
\end{eqnarray}

For $p>0$ and $t>a$ we obtain
\begin{eqnarray}\label{important}
{}_a \textbf{D}_t^{p}\left({}_a \textbf{D}_t^{-p}f(t)\right)=f(t),
\end{eqnarray}
which means that the RL fractional differentiation operator is a
left inverse to the RL fractional integration operator of the same
order. The relation (\ref{important}) is  called the fundamental
property of the RL fractional derivative. In addition, the
fractional derivative of a constant is  not zero and the RL
fractional derivative of the power function $(t-a)^\nu$ is given
by
\begin{eqnarray}
{}_a
D_t^{p}(t-a)^\nu=\frac{\Gamma{(\nu+1)}}{\Gamma{(-p+\nu+1)}}(t-a)^{\nu-p},
\end{eqnarray}
where $\nu >-1$. The normal derivatives $\frac{d^n}{dt^n}$ and
${}_a D_t^{p}$ commute only if $ \ f^{(j)}(a)=0, \ j=0,1,\dots,n-1
\ $ is fulfilled and  two RL fractional derivative operators ${}_a
\textbf{D}_t^{p}$ and ${}_a \textbf{D}_t^{q}$ commute only if
\begin{eqnarray}
\left[{}_a \textbf{D}_t^{p-j}f(t)\right]_{t=a} & = & 0 \ ,\qquad
j=1,\cdots,m
\end{eqnarray}
and
\begin{eqnarray}
\left[{}_a \textbf{D}_t^{q-j}f(t)\right]_{t=a}  & = & 0 \ ,\qquad
j=1,\cdots,n \ .
\end{eqnarray}
The above properties of RL fractional derivatives lead us to the
conclusion that  there are many substantial differences from the
usual derivatives and therefore the solutions of the differential
fractional equations contain more information than the classical
ones.

\section{Fractional Euler-Lagrange equations for\\
 Lagrangians with linear velocities}

Let $J[q^1,\cdots,q^n]$ be a functional of the form
\begin{eqnarray}
\int\limits_a^b L\left(t,q^1,\cdots,q^n,{}_a
\textbf{D}_t^{\alpha}q^1,\cdots,{}_a\textbf{D}_t^{\alpha}q^n,{}_t
\textbf{D}_b^{\beta}q^1,\cdots,{}_t
\textbf{D}_b^{\beta}q^n\right)dt
\end{eqnarray}
defined on the set of functions $q^i(t), \ i=1,\cdots,n$ which
have continuous left RL fractional derivative of order $\alpha$
and right RL fractional derivative of order $\beta$ in $[a,b]$ and
satisfy the boundary conditions $q^i(a)=q^i_a$ and $q^i(b)=q^i_b$.
 A necessary condition for $J[q^1,\cdots,q^n]$ to admit an
extremum for  given functions $q^i(t), \ i=1,\cdots, n$ is that
$q^i(t)$ satisfy  Euler-Lagrange equations \cite{om1}
\begin{eqnarray}\label{lola}
\frac{\partial L}{\partial q^j}+{}_t
\textbf{D}_b^{\alpha}\frac{\partial L}{\partial {}_a
\textbf{D}_t^{\alpha} q^j}+{}_a \textbf{D}_t^{\beta}\frac{\partial
L}{\partial {}_t \textbf{D}_b^{\beta} q^j}=0 \ ,\qquad
j=1,\cdots,n \ .
\end{eqnarray}
 In the following we consider the Lagrangian with linear
 velocities
\begin{equation}\label{lagra1} L=a_{j}\left(q^{i}\right)\dot
q^{j}-V\left(q^{i}\right),
\end{equation}
where $a_{j}\left(q^{i}\right)$ and $V\left(q^{i}\right)$ are
functions of their arguments.

The fist step is to construct the corresponding fractional
generalization of the Lagrangian given by (\ref{lagra1}).  The
fractional Lagrangian is not unique, in other words there are
several possibilities to replace the time derivative with
fractional ones. The requirement is to obtain the same Lagrangian
expression if the order $\alpha$ is 1. Having in mind the above
considerations, for $0<\alpha\leq 1$, we propose two fractional
Lagrangians. The first one is as follows

\begin{equation}\label{lagra2}
L'=a_{j}\left(q^{i}\right){}_a \textbf{D}_t^{\alpha}
q^{j}-V\left(q^{i}\right) \ .
\end{equation}
From (\ref{lola}) and (\ref{lagra2}), the corresponding
Euler-Lagrange equations emerge as
\begin{equation}\label{el1}
\frac{\partial a_{j}(q^i)}{\partial q^{k}}{}_a
\textbf{D}_t^{\alpha} q^{j}+{}_t
\textbf{D}_b^{\alpha}a_{k}\left(q^{i}\right)-\frac{\partial
V\left(q^{i}\right)}{\partial q^{k}}=0 \ .
\end{equation}
The second Lagrangian is given by
\begin{equation}\label{lagra3}
L'=-a_{j}\left(q^{i}\right){}_t \textbf{D}_b^{\alpha}
q^{j}-V\left(q^{i}\right) \ .
\end{equation}
Using (\ref{lola}) and (\ref{lagra3}) the corresponding
Euler-Lagrange equations become
\begin{equation}\label{el2}
\frac{\partial a_{j}(q^i)}{\partial q^{k}}{}_t
\textbf{D}_b^{\alpha} q^{j}+{}_a
\textbf{D}_t^{\alpha}a_{k}\left(q^{i}\right)+\frac{\partial
V\left(q^{i}\right)}{\partial q^{k}}=0 \ .
\end{equation}
\subsection{Examples}

       {\bf A.}
       To illustrate our approach, let us consider the following
       Lagrangian
\begin{equation}\label{lala}
L=\dot q^1 q^2-\dot q^2 q^1-(q^1-q^2) q^3
\end{equation}
which is a gauge invariant \cite{example1}. In this case we
proposed  the corresponding fractional Lagrangian to be as

\begin{equation}\label{lala1}
L'=\left({}_a \textbf{D}_t^{\alpha} q^1\right) q^2-\left({}_a
\textbf{D}_t^{\alpha} q^2\right) q^1-(q^1-q^2) q^3 \ .
\end{equation}

Using (\ref{el1}), the Euler-Lagrange equations corresponding to
(\ref{lala1}) become

\begin{equation}\label{equation}
 q^1=q^2 \ , \quad
  -{}_a
\textbf{D}_t^{\alpha}q^2-q^3+{}_t \textbf{D}_b^{\alpha}q^2=0 \ ,
\quad
 {}_a \textbf{D}_t^{\alpha}q^1+q^3-{}_t
\textbf{D}_b^{\alpha}q^1=0 .\
\end{equation}

 The solution of (\ref{equation}) is given as follows
\begin{equation}\label{q1}
q^1=q^2 \ ,
\end{equation}

\begin{equation}\label{q3}
q^3=  (-{}_a \textbf{D}_t^{\alpha} +
{}_t\textbf{D}_b^{\alpha})q^1.
\end{equation}

From (\ref{q1}) and (\ref{q3}) we conclude that the classical
solution is obtained
if $\alpha\rightarrow{1}$.\\

 {\bf B.} Let us consider the second  Lagrangian  given by

  \begin{equation}\label{lals}
  L=\dot q^1 q^2+\dot q^3 q^4- V(q^2,q^3,q^4),
  \end{equation}
  where  $V(q^2,q^3,q^4)= {-1\over 2}[(q^4)^2-2q^2q^3]$.  We observe that (\ref{lals}) is a  second class
  constrained system  in Dirac's classification \cite{const}.

  We propose the fractional generalization of (\ref{lals}) to be
  as follows

 \begin{equation}\label{lf}
L'=-[({}_t \textbf{D}_b^{\alpha}q^1)q^2+({}_t
\textbf{D}_b^{\alpha}q^3)q^4+V(q^2,q^3,q^4)] \ .
   \end{equation}

   From  (\ref{el2}) and (\ref{lf}) the Euler-Lagrange equations are given by

\begin{equation}\label{ecuati1}
     {}_a \textbf{D}_t^{\alpha}q^2=0 \ ,
      \end{equation}
      \begin{equation}\label{ecuati2}
     {}_t\textbf{D}_b^{\alpha}q^1+q^3=0 \ ,
     \end{equation}
      \begin{equation}\label{ecuati3}
    {}_a\textbf{D}_t^{\alpha}q^4+q^2=0 \ ,
       \end{equation}
\begin{equation}\label{ecuati4}
     {}_t\textbf{D}_b^{\alpha}q^3-q^4=0 \ .
     \end{equation}

From (\ref{ecuati1}), we conclude that  the solution for $q^2(t)$
has the form
\begin{equation}\label{eqq2}
     q^2(t)=C_1(t-a)^{\alpha-1} \ .
       \end{equation}

From (\ref{ecuati3}) and (\ref{eqq2}) an equation for $q^4(t)$ is
obtain as follows
 \begin{equation}\label{eqq3}
    {}_a\textbf{D}_t^{\alpha}q^4=-C_1(t-a)^{\alpha-1}.
       \end{equation}

The solution of (\ref{eqq3}) has the form

\begin{equation}\label{ecu4}
q^4(t)=C_2(t-a)^{\alpha-1}-\frac{C_1}{\Gamma{(\alpha)}}\int\limits_a^t(-\tau+t)^{\alpha-1}(\tau-a)^{\alpha-1}d\tau
\ .
\end{equation}

Using (\ref{ecuati4}) and (\ref{ecu4}), the solution of $q^3(t)$
becomes

\begin{eqnarray}\label{eee}
     q^3(t) & = &
     C_3(b-t)^{\alpha-1}+\frac{C_2}{\Gamma{(\alpha)}}\int\limits_t^b(-t+\tau)^{\alpha-1}(-a+\tau)^{\alpha-1}d\tau\nonumber\\
     & {} &
     -\frac{C_1}{{\Gamma{(\alpha)}}^2}\int\limits_t^b(-t+\tau)^{\alpha-1}\int\limits_a^\tau(\tau-\eta)^{\alpha-1}(\eta-a)^{\alpha-1}d\eta
     d\tau \ .
\end{eqnarray}

Finally, the equation (\ref{ecuati2}) together with (\ref{eee})
give the solution for $q^1(t)$ as follows

\begin{eqnarray}\label{ehe}
     q^1(t) & = & C_4(b-t)^{\alpha-1}-\frac{C_3}{\Gamma{(\alpha)}}\int\limits_t^b(-t+\tau)^{\alpha-1}(b-\tau)^{\alpha-1}d\tau\nonumber\\
     & {} & -\frac{C_2}{\Gamma{(\alpha)}^2}\int\limits_t^b(-t+\tau)^{\alpha-1}\int\limits_\tau^b(-\tau+\eta)^{\alpha-1}(-a+\eta)^{\alpha-1}d\eta d\tau\nonumber\\
     & {} & +\frac{C_1}{\Gamma{(\alpha)}^3}\int\limits_t^b (\sigma-t)^{\alpha-1}\int\limits_\sigma^b(\sigma-\eta)^{\alpha-1}\int\limits_a^\eta(\tau-\eta)^{\alpha-1}(\eta-a)^{\alpha-1}d\tau
     d\eta d\sigma \ .
\end{eqnarray}
Here $C_1,C_2$, $C_3$ and $C_4$ are constants. If
$\alpha\rightarrow 1$, $a\rightarrow 0$, $b\rightarrow 1$, then
the standard solutions \be q^1(t)={C^{'}_4 t^3\over 6}-
{C^{'}_3t^2\over 2}+C^{'}_2t +C^{'}_1, \ q^2(t)=C^{'}_4, \
q^3(t)={C^{'}_4\over 2}t^2-C^{'}_3t+C^{'}_2, \
q^4(t)=-C^{'}_4t+C^{'}_3 \ee are recovered if we redefine the
constants from (\ref{eqq2}), (\ref{ecu4}), (\ref{eee}) and
(\ref{ehe}) as

 $\ C_1^{'}=C_4 -C_3-{{C_2\over 2}} +{C_1\over 3},\ C_2^{'}={-C_1\over 2}+C_2 +C_3,\ C^{'}_3=C_2,C^{'}_4=C_1$.

  \section{Conclusion}

  The Lagrangians with linear velocities were investigated using
  left and right RL fractional derivatives. The corresponding fractional
Lagrangians were proposed  and the fractional Euler-Lagrange
equations were obtained. Although the fractional Lagrangians
contain only left or right derivatives, both derivatives are
involved in Euler- Lagrange equations and both played an important
role in finding the solutions.
 The exact solutions of the Euler-Lagrange equations
  were obtained for two examples corresponding to first and second-class constrained systems.
 A ``gauge invariance'' was reported for the first example.
The solutions of the investigated examples depend on the limits a
and b and the limiting procedure recovered the standard results.

 \section {Acknowledgments}
This work is partially supported by the Scientific and Technical
Research Council of Turkey. One of the authors (D. B.) would like
to thank R. Hilfer, O. Agrawal and N. Engheta for providing him
important references and M. Naber and M. Henneaux for interesting
discussions.

\end{document}